\documentclass[12pt]{article}

\newcommand{\aprox}{approximation}
\newcommand{\be}{\begin{equation}}
\newcommand{\bm}{Bohmian motion}
\newcommand{\BM}{Bohmian mechanics}

\renewcommand{\c}{classical}
\newcommand{\ca}{classical action}
\newcommand{\cb}{classical behavior}
\newcommand{\com}{center of mass}

\newcommand{\conv}{convergence}
\newcommand{\cm}{classical motion}
\newcommand{\CM}{classical mechanics}
\newcommand{\cl}{classical limit}

\newcommand{\cw}{classical world}

\renewcommand{\d}{\delta}

\newcommand{\db}{de Broglie}
\newcommand{\deco}{decoherence}

\newcommand{\dof}{degrees of freedom}
\newcommand{\e}{\epsilon}
\newcommand{\ee}{\end{equation}}

\newcommand{\Ehrt}{Ehrenfest theorem}
\newcommand{\env}{environment}

\newcommand{\ex}{\textrm{e}}

\newcommand{\extp}{external potential}

\renewcommand{\Im}{\mbox{Im}}

\newcommand{\h}{\hbar}

\newcommand{\HJ}{Hamilton-Jacobi}

\renewcommand{\l}{\lambda}

\newcommand{\LPW}{local plane wave}
\newcommand{\LPWs}{local plane waves}
\newcommand{\lpwst}{local plane wave structure}

\newcommand{\ode}[2]{\frac{ d #1}{ d #2}}
\newcommand{\pd}{probability distribution}
\newcommand{\pde}[2]{\frac{\partial #1}{\partial #2}}

\newcommand{\psie}{\psi^\epsilon}

\newcommand{\QM}{quantum mechanics}
\newcommand{\qp}{quantum potential}

\renewcommand{\r}{\rho}

\newcommand{\ri}{\rightarrow}
\newcommand{\RR}{I\!\!R}

\newcommand{\se}{Schr\"odinger's equation}

\newcommand{\sev}{Schr\"odinger's evolution}

\newcommand{\sv}{slowly varying}
\renewcommand{\t}{\tau}

\newcommand{\tr}{trajectory}

\newcommand{\vel}{velocity}
\newcommand{\virt}{``virtual''}
\newcommand{\vps}{wave packets}
\newcommand{\vp}{wave packet}
\newcommand{\wf}{wave function}
\newcommand{\wfs}{wave functions}
\newcommand{\wl}{wave length}
\newcommand{\Xe}{X^{\epsilon}}

\begin{document}

\title{On the Classical Limit of Quantum Mechanics}
\author{Valia Allori \footnote{e-mail: allori@ge.infn.it}
\and Nino Zangh\`{\i} \footnote{e-mail: zanghi@ge.infn.it}  }
\date{\small Dipartimento di Fisica dell'Universit\`a di Genova\\
Istituto Nazionale di Fisica Nucleare, Sezione di Genova\\
via Dodecaneso 33, 16146 Genova, Italy}
\maketitle

\begin{abstract}
Contrary to the widespread belief, the problem of the emergence of
classical mechanics from quantum mechanics is still open.  In spite of
many results on the $\h\ri 0$ asymptotics, it is not yet clear how to
explain within standard \QM{} the \cm{} of macroscopic bodies.  In
this paper we shall analyze special cases of classical behavior in the
framework of a precise formulation of \QM{}, \BM{}, which contains in
its own structure the possibility of describing real objects in an
observer-independent way.
\end{abstract}

\section{Introduction}

According to the general wisdom there shouldn't be any problem with
the \cl{} of \QM{}.  In fact, in any textbook of \QM{} one can easily
find a section where the solution of this problem is explained (see,
e.g., \cite{schiff}) through \Ehrt{}, WKB \aprox{} or simply the
observation that the canonical commutation relations become Poisson
brackets.  Indeed, one might easily get the impression that it is only
a matter of putting all known results into order for obtaining a
rigorous derivation of classical mechanics {}from \QM{}.  Consider,
for example, the standard argument based on the \Ehrt{}: if the
initial \wf{} is a narrow \vp{}, the packet moves approximately
according to  Newtonian law $F=ma$.  This argument entails
that classicality is somehow associated with the formation and
preservation of narrow wave packets.

There are, however, several problems connected with this.  The most
serious one is that a \vp{} typically spreads and there is a definite
time after which the \c{} \aprox{} will break down.  Even if it might
appear that, for massive bodies, the \vp{} will remain narrow for very
long times, it can easily be shown that interactions will typically
generate very spread out \wfs{}, even for massive bodies (e.g, for the
center of mass of an asteroid undergoing chaotic motion).  But which
mechanism should prevent the wave function from spreading?

Recently, it has been suggested that {\it decoherence}, due to the
interaction of the system with its environment could provide the
desired mechanism: The \env{}, constantly interacting with the body,
could somehow act as a measuring device of the macroscopic variables
of the body---say, its center of mass---producing in this way a narrow
wave function in the macroscopic directions of its configuration space
(see \cite{libro} and references therein).  But the composite system
formed by the system of interest and its \env{} is itself a closed
system and Schr\"odinger's evolution of this enlarged system tends to
produce spreading of the total wave function over the total
configuration space.  Thus, \deco{} alone is not sufficient to explain
the emergence of the \cw{} from standard \QM{}: there is still the
necessity to add to \se{} the collapse of the wave function.  Within
standard quantum mechanics, this is the only way to guarantee the
formation of a narrow wave function in the macroscopic directions of
the configuration space of the system.  However, Schr\"odinger's
evolution plus the collapse is not a precise microscopic theory: the
division between microscopic and macroscopic world (where the collapse
takes place) is not part of the theory.  Thus, as Bell has suggested
\cite{bell}, we need to go beyond standard \QM{}: or the \wf{} doesn't
provide the whole description of the state of a system or \se{}
needs to be modified (for a proposal of this kind, see \cite{grw} ).
\medskip

In this paper we shall move the first steps towards a complete
derivation of the \cl{} within the framework of \BM{}, a theory about
particles moving in physical space and which accounts for all quantum
phenomena.  While we refer to future work for a complete and thorough
analysis (see \cite{allori0} and \cite{allori2}) we shall show here
how \BM{} allows to go much beyond the standard approach and to
explain the emergence of classicality even for spread out wave
functions.  We shall formulate the \cl{} as a scaling limit in terms
of an adimensional parameter $\e$ given by the ratio between two
relevant length scales, namely, the ``wave length'' of the particle,
and the ``scale of variation'' of the potential $L$.  We shall show
how, in some special cases, the emergence of classical behaviour is
associated with the limit $\e\to 0$.

\section{Classical Limit in Bohmian Mechanics}

\BM{} is a theory in which the world is described by particles whose
configurations follow trajectories determined by a law of motion.  The
state of a system is described by the couple $(X,\psi)$, where
$X=(X_1,...,X_N)$ are the configurations of the particles composing
the system and $\psi$ is the \wf{} evolving according to \se{}.  For a
review of \BM{}, see the other contribution to the same volume and
references therein \cite{allori1}.

In order to formulate the \cl{} in \BM{}, it can be useful to write
the \wf{} $\psi$ in the polar form $\psi=R\ex^{\frac{i}{\h}S}$.
{}From \se{}
$$
i\hbar\pde{\psi}{t} = -\sum_{k=1}^N\frac{\h^2}{2m_k}\nabla_k^2 \psi
+V\psi \,,
$$
one obtains the continuity equation for $R^2$, \be
\frac{\partial R^2}{\partial t}+{\rm{div}}\left[\left( \frac{\nabla_k
S}{m}\right)R^2\right]=0,
\label{eq:qcont}
\ee and the modified \HJ{} equation for $S$ \be \frac{\partial
S}{\partial t}+\frac{(\nabla_k S)^2}{2m}+
V-\sum_k\frac{\h^2}{2m_k}\frac{\nabla_k^2 R}{R}=0.
\label{eq:qHJ}
\ee Equation (\ref{eq:qcont}) suggests that $\r=R^2$ can be
interpreted as a probability density.  Note that equation
(\ref{eq:qHJ}) is the usual classical \HJ{} equation with an
additional term \be U\equiv-\sum_k\frac{\h^2}{2m_k}\frac{\nabla_k^2
R}{R},
 \label{eq:potential}
\ee called the quantum potential.  One then sees that the (size of
the) quantum potential provides a rough measure of the deviation of
\BM\ from its classical approximation.

In this way, it might seem that the \cl{} is something trivial: ensure
the \qp{} is somehow small and then \CM{} arises {}from \BM{}.  What
is not trivial at all is to understand what are the physical
conditions corresponding to the smallness of the \qp{}.  In the next
sections we shall discuss a simple model of a macroscopic body moving
in an \extp{} and we shall see that there exists a precise limit in
which the time evolution of the center of mass of the body is
approximately \c{}.

\subsection{Motion in an External Potential}
\label{subsec:Motion in an External Potential}

Consider a macroscopic body composed of $N$ particles, with positions
$(X_1,...,X_N)$, and masses $m_1, \ldots, m_N$, subjected only to
internal forces.  The
\com{} of the body is $$X=\frac{\sum_i m_i X_i}{\sum_i m_i}. $$
The Hamiltonian of the body can be written as \be
H=\frac{\hbar^2}{2m}{\nabla}^2+H_{\rm rel}, \label{eq:free} \ee where
$m\equiv \sum_i m_i$, and $H_{rel}$ is depending only on the relative
coordinates $y_i=x_i-x$ (and their derivatives).  Therefore, if the
wave function of the body $\Psi=\Psi(x,y)$ (at some ``initial'' time)
factorizes as a product of the wave function $\psi=\psi(x) $ of the
\com{} and the \wf{} $\phi=\phi(y)$ of the internal \dof{}, this
product form will be preserved by the dynamics and the state
$(X,\psi)$ will evolve autonomously: $\psi$ satisfies to \se{} with
Hamiltonian (\ref{eq:free}) and $X$ evolves according to \be
\frac{dX_t}{dt}=\frac{\h}{m}{\rm Im} \left[\frac{\nabla
\psi(X_t,t)}{\psi(X_t,t)}\right].
\label{eq:bohm}
\ee

If there is an
\extp{} $ \sum_i V_i(q_i)$, the Hamiltonian will assume the form
\be
H=\frac{\hbar^2}{2m}{\nabla}^2 +V(x) + H^{(x,y)}
\label{eq:hamex}
\ee where $V(x) = \sum_i V_i(x)$, $H^{(x,y)} =H_{\rm rel}+H_{\rm
int}$, with $H_{\rm int}$ describing the interaction between the
\com{} and the relative coordinates.  If $V_{i}$ are \sv{} on the size
of the body, $H_{\rm int}$ can be treated as a small perturbation, and,
in first approximation, neglected.  Thus, if
$\Psi=\Psi(x,y)=\psi(x)\phi(y)$, at some time, this product form will
be preserved by the dynamics.  In this way we end up again with a very
simple one body problem: $\psi$ evolves according to \se{} with
Hamiltonian $$ H=\frac{\hbar^2}{2m}{\nabla}^2 +V(x)
$$
and the position $X$ of the center of mass of the body evolves
according to (\ref{eq:bohm}).

\subsection{Classical Limit as a Scaling Limit}
\label{subsec:Conjecture on the Emergence of the Classicality}

Usually, the classical limit is associated with the limit $\h\ri 0$,
meaning by this $\h\ll A_0$, where $A_0$ is {\it some} characteristic
action of the corresponding \cm{} (see, e.g., \cite{maslov},
\cite{schiff}, \cite{berry}).  Note that, while $\psi$ doesn't have
any limit as $\h$ goes to zero, the couple $(R,S)$, defined by the
$\h$-dependent change of variables $\psi=R\ex^{\frac{i}{\h}S}$, does
have a limit.  Formally, this limit can be read by setting $\h\equiv0$
in equations (\ref{eq:qcont}) and (\ref{eq:qHJ}) and thus the couple
$(R,S)$ becomes the pair $(R^0,S^0)$, where $R^0$ satisfies the \c{}
continuity equation and $S^0$ the \c{} \HJ{} equation.

The condition $\h\ll A_0$ is often regarded as equivalent to another
standard condition of classicality which involves some relevant length
scales of the motion: if the \db{} \wl{} $\l$ is small with respect to
the characteristic dimension $L$ determined by the scale of variation
of the potential, the behavior of the system should be close to the
\cb{} in the same potential (see, e.g., \cite{landau}).  This
condition relates in a completely transparent way a property of the
state, namely its \wl{}, and a properties of the dynamics, namely the
scale of variation of the potential.

We shall not enter here into the problem of finding a precise
characterization of $\l$ and $L$ (for which we refer to \cite{allori0}
and \cite{allori2}).  For the present purposes it is sufficient to
keep in mind that the \db{} \wl{} should be regarded as function of
the initial \wf{}, $\l=\l(\psi_0)$, e.g., given in terms of the mean
kinetic energy with respect to $\psi_0$, and the scale of variation of
the potential should be regarded as a suitable function of the
potential, $L=L(V)$, for example, for a potential of the form
$V(x)={\rm sin}(\frac{2\pi}{a} x)$, $L=a$, for a constant potential
(free case) $L=\infty$.

The length scales $\l$ and $L$ allow to define the natural macroscopic
scales $x'$ and $t'$ for describing the motion, \be x' =
\frac{x}{L}\,,\qquad t'=\frac{t}{T}.
\label{eq:macroscales}
\ee The time scale $T$ is defined as $T=\frac{L}{v}$ and $v$ is the
speed defined by $\l$, $v=\frac{\h}{m\l}$.  The scales $L$ and $T$
tell us what are the fundamental units of measure for the motion:
$L$ is the scale in which the potential varies and $T$ is the time
necessary to the particle to see its effects.  We expect the \bm{} on
these scales to look \c{} when the adimensional parameter $$\e\equiv
\frac{\l}{L}$$ is getting smaller and smaller.  This means in
particular that we expect the \qp{} to become negligible (with respect
to the kinetic energy) whenever $\e\ri 0$.

In the next two sections we shall study some examples of wave functions
and potentials for  which we can show explicitly that the \cl{} arises.

\section{Quasi Classical Wave Functions}
\label{subsec:Quasi Classical Wave Functions}

Consider a family of \wfs{} depending on $\h$ of the {\it short wave}
form \be \psi_0^{\h}(x)=R_0(x)\ex^{\frac{i}{\h}S_0(x)},
\label{eq:shortwf0}
\ee where $R_0(x)$ and $S_0(x)$ are functions not depending on $\h$,
and $R(x)$ is not zero only in a limited region of space.  The limit
$\h\ri 0$ corresponds to a mathematical trick to simulate the limit
$\l(\psi_0)\ri 0$.  This limit, called short \wl{} limit, is a special
case of the limit $\e=\l/L\ri 0$ in which the \wl{} is getting small
and the \extp{} is fixed.  To see the \cl{} arising, one should
describe the motion in terms of the macroscopic coordinates $x' =x/L$
and $t'=t/T$.  However, since in this case both $L$ and $T$ are
constant, there is no substantial difference between the microscopic
and macroscopic scale.  (Note that the limit $\h\ri 0$ simulates also
the limit of large mass $m\ri +\infty$ and for which the potential
rescales as $V=m\hat{V}$.  In
fact, in this case, $1/m$ and $\h$ play the same role in the \se{} $$
i\frac{\partial \psi}{\partial t}=\left[
-\frac{\h}{2m}\nabla^2+\frac{m}{\h}\hat{V}\right]\psi.
$$
and in the guiding equation (\ref{eq:bohm}).)

The approximate solution of \se{} for initial condition
(\ref{eq:shortwf0}) in the short \wl{} limit is given by
\footnote{\label{foot1} We are considering here times shorter than the
``first caustic time'', i.e. the time at which the function
$p(x,t)=\nabla S^0(x,t)$ becomes multivalued.  This in not a
restriction, as explained in \cite{allori0} and \cite{allori2}.  }
\cite{maslov} \be
\psi^{(0)}(x,t)=R^{(0)}(x,t)\ex^{\frac{i}{\h}S^{(0)}(x,t)}+O(\h),
\label{eq:limitshortwf}
\ee where $S^{(0)}(x,t)$ is the \ca{} (i.e., the solution of \c{}
\HJ{} equation having $S_0(x)$ as initial condition) and
$R^{(0)}(x,t)=|dx/dx_0|^{-1/2}R_0(x_0,t)$ (where $x_0=x_0(x,t)$ is the
initial position of the particle that at time $t$ arrives in $x$
transported along the classical path).
$R^{(0)}(x,t)$ is the evolution at time $t$ of the initial amplitude
$R_0(x,t)$ according to the \c{} continuity equation (\ref{eq:qcont}).

Therefore, in the limit $\h\ri 0$, the \vel{} field becomes the \c{}
one \be v^{(0)}(x,t)=\frac{1}{m}\nabla S^{(0)}(x,t)+O(\h^2).
\label{eq:limithvelofield}
\ee and the \qp{} \be U=-\frac{\h^2}{m}\frac{\nabla^2
{R^{(0)}}}{R^{(0)}} \ee goes zero because $R^{(0)}$ doesn't depend on
$\h$.  In other words, in the limit $\h\ri 0$, we have \conv{} of
modified \HJ{} equation to \c{} \HJ{} equation.

\section{Slowly Varying Potentials}
\label{subsec:Slowly Varying Potentials}

Another special case of the limit $\e\ri 0$ is given by the situation
in which there is a \sv{} \extp{}, $$V_{L}=V\left(\frac{x}{L}\right),$$
with scale of variation $L$ very large.  Given that $\e=\l/L$, the
\sv{} potential limit is a special limit corresponding to keep $\l$
fixed (that is, the initial \wf{}) and letting $L\ri +\infty$.  This
limit is equivalent to a long time limit.  In fact, if the potential
is \sv{} ($L\ri +\infty$), to see its effect the particle has to
travel
for a time of order $$T=\frac{mL\l}{\h} ,$$
and for $L\to\infty$ also $T\to\infty$.

Since time and space rescaling are of the same order, it is useful to
rescale space and time as \be x\ri \frac{x}{\e},\quad t\ri
\frac{t}{\e}.  \ee Under these rescaling, the initial conditions of
Bohmian dynamics become \be
\psie_0(x)=\frac{1}{\e^{1/2}}\psi_0(\frac{x}{\e}), \quad
X_0^{\e}=\frac{X_0}{\e} \ee and Bohmian equations become the usual
equations of motion for $\psie(x,t)$ with $\h$ substituted by $\h\e$
\begin{eqnarray}
  i\hbar\e\pde{\psie(x,t)}{t}&=& -\frac{\h^2\e^2}{2m} \nabla^2\psie(x,t)
  +V(x)\psie(x,t),
 \label{eq:scepsilon}\\
 \ode{\Xe_t}{t}& =& v^{\e}(\Xe_t,t)= \frac{\h\e}{m}\Im
  \left[\frac{\nabla\psie(\Xe_t,t)}{\psie(\Xe_t,t)}\right].
 \label{eq:bohmepsilon}
\end{eqnarray}

The solution $\psie(x,t)$ of equation (\ref{eq:scepsilon}) can be
expressed in terms of the rescaled propagator $K^{\e}(x,t;x_0,0)$, in
which $\h$ is replaced by $\h\e$, and the Fourier transform of the
initial \wf{} $\hat{\psie}_0(k)$, \be
\psie(x,t)=\frac{1}{(2\pi\e)^{d/2}}\int\int K^{\e}(x, t; x_0,0)
\ex^{i\frac{{x_{0}}\cdot{k}}{\e}} \hat{\psi_0}(k) d^dx_0 d^dk ,
\label{eq:psie}
\ee where {\it d} is the dimension of the space.  In general, the
asymptotic form of the propagator in the limit $\e\ri 0$ is
\cite{gutzwiller} \be K^{\e}(x,t;x_0,0)=\frac{1}{(2\pi
i\h\e)^{d/2}}\sqrt{C(x,x_0;t)} \ex^{\frac{i}{\h\e} {S^{0}}(x,t;x_0,0)}
[1+\h\e z],
\label{eq:KeG}
\ee where $z=z(t,x_0,x,\e)$ and $||z||_{L^2({\bf \RR^d})}\le c$ where
$c$ is a constant \footnote{We are again assuming times shorter than
the ``first caustic time''.  See footnote (\ref{foot1}).}.  $S^0$ is
the \ca{} \be S^0(x,t)=\int_0^{\t}L(x(\t),\dot{x}(\t),\t)d\t,\qquad
x(0)=x_0,\quad x(t)=x, \ee and \be C(x,x_0;t) ={\rm
det}\left[-\nabla_{x,{x_{0}}}^2 S^0(x,t;x_0,0) \right].  \ee

In order to find the asymptotic $\e\ri 0$ of $\psie(x,t)$, we apply
the method of stationary phase: the main contribution to $\psie(x,t)$
comes {}from the $x_0$ and the $k$ which make stationary the phase
$\phi(x_0,k)=\frac{1}{\h}[S^0(x,t;x_0,0)+x_0\cdot\h k]$.  They are \be
x_0=0 \quad\textrm {and}\quad k_0(x,t)=-
\frac{1}{\h}\left.\nabla_{x_0}S^0(x,t;x_0,0) \right|_{{x_{0}}=0}.
 \label{eq:stationary}
  \ee So we have \be \psi^{(0)}(x,t)= \sqrt{C(x,0;t)}
  \left(\frac{i}{\h}\right)^{d/2}
  \hat{\psi}_0(k_0(x,t))\ex^{\frac{i}{\h\e} S^0(x,t;0,0)}+O(\e).
  \label{eq:limitpsie}
 \ee
We can rewrite this as
\be
   \psi^{(0)}(x,t)=R^0(x,t)\ex^{\frac{i}{\h\e}S^0(x,t)}+O(\e^2),
\label{eq:psielpw}
    \ee where 
\be
R^0(x,t)=\sqrt{C(x,0;t)}\left(\frac{i}{\h}\right)^{d/2}\hat{\psi}_0(k_
0(x,t)).
\label{eq:limitRe}
\ee
In the limit $\e \ri 0$, the \vel{} field becomes \be
v^{(0)}(x,t)=\frac{1}{m}\left.\nabla_{x}
S^0(x,t;x_0,0)\right|_{x_0=0}+O(\e^2).
 \label{eq:limitvelofielde}
\ee

Therefore, also in this case, we have \conv{} of modified \HJ{} equation
to the \c{} \HJ{} equation (and thus vanishing of the quantum
potential, as before) although for different initial conditions.  In
fact, in the case of the family of quasi classical wave functions ($\h\ri
0$ limit), the limiting position is distributed according to the \c{}
\pd{} $\r(x,t)=|R^0(x)|^2$ and the limiting \vel{} is distributed
according to \be \r(v,t)=\int \d\left(v-\frac{\nabla S^0(x,t)}{m}
\right)\left|R^0(x)\right|^2dx.  \ee On the other hand, in the case of
\sv{} potential, the \pd{} of the limiting position is \be
\r(x,t)=\frac{C(x,0;t)}{\h^d} |\hat{\psi}_0(k_0(x,t))|^2,
\label{eq:pdxe}
\ee where $k_0(x,t)$ is defined by equation (\ref{eq:stationary}).
This is the \pd{} transported along the \c{} flow for the initial
position is $X_0=0$ and  initial velocity  distributed
according to \be \r(v,0)=\left(\frac{m}{\h}\right)^d
\left|\hat{\psi}_0\left(\frac{mv}{\h}\right)\right|^2.
\label{eq:pdve}
\ee
Note
the presence of the Fourier transform of the initial \wf{} in the
\pd{} of the initial velocity, which is somehow connected to the fact
that the \sv{} potential limit is equivalent to a long time limit in
which the initial transient behaviour is completely canceled out.

\section{General Structure of the Classical Limit}
\label{sec:Local Plane Wave Structure}

The examples discussed in the previous sections show that there is a
particular structure of the \wf{} (see equations
(\ref{eq:limitshortwf}) and (\ref{eq:psielpw})) that emerges when we
are in the \c{} regime.  This structure is what we call a {\it local
plane wave}, a \wf{} that locally can be regarded as a plane wave
having a local \wl{}.

A precise notion of \LPW{} can be given starting {}from the usual
notion of \wl{} $\l$ (the spatial period).  For simplicity, we shall
analyze the problem in one dimension.  Consider a \wf{} of the polar
form, then the following relations should hold
\begin{eqnarray}
R(x,t)&\simeq& R(x+\l,t),\label{eq:condR}      \\
S(x,t)&\simeq& S(x+\l,t)+2\pi\h.\label{eq:condS}
\end{eqnarray}
By expanding in Taylor series in $\l$ the right hand side of equation
(\ref{eq:condR}), one gets
\begin{eqnarray}
\left|\frac{\nabla R(x,t)}{R(x,t)}\right|\l(x,t)&\ll&
1,\label{eq:condR1}\\
\frac{1}{2}\left|\frac{\nabla^2 R(x,t)}{R(x,t)}\right|\l^2(x,t)&\ll&
1,\quad ...  \label{eq:condR2}
\end{eqnarray}
Similarly for $S(x,t)$ we obtain, up to the second order terms, the
definition of the local \wl{} $\l(x,t)$ \be \l(x,t)=\frac{\h}{|\nabla
S(x,t)|}.
\label{eq:deflambdaloc}
\ee The smallness of the second order term, together with equation
(\ref{eq:deflambdaloc}), gives the condition \be
\left|\nabla\l(x,t)\right|\ll 1.
\label{eq:condl}
\ee

It should be stressed that condition (\ref{eq:condR2}) directly
implies that the \qp{} is smaller than the kinetic energy for a given
time $t$, i.e. \be \frac{\h^2}{2m}\left|\frac{\nabla^2
R(x,t)}{R(x,t)}\right| \ll \frac{1}{2m}\left|\nabla S(x,t)\right|^2,
\label{eq:qpke}
\ee which, in its turn, implies the validity of the \c{} \HJ{}
equation.  We may then conclude that the association between the
emergence of \cb{} and the formation of \LPWs{} is indeed the hallmark
of the \cl{}.  This conclusion receives further supports {}from
observing the expansive character of the Laplacian in \se{} which
tends to produce spreading of the \wf{} insofar the potential energy
is dominated by the kinetic energy (that is, for bounded motion in a
potential far {}from the turning points).

Moreover, observe that to have \cl{}, the quantum potential should be
smaller than the kinetic energy for a sufficiently large time
interval.  In other words, classicality requires that the \lpwst{}
should be preserved by the dynamics.  There is an argument, based on
Ehrenfest theorem, that allows to determine the correct notion of
scale of variation of the potential $L$ and explains the the stability
of the \lpwst{} in terms of the condition $\l\ll L$.  It is outside
the scope of this paper to enter into the details of this argument
(see \cite{allori0} and \cite{allori2}).  It turns out that $L$ is
given by \be L=\sqrt{\frac{|V'(x)|}{|V'''(x)|}},
\label{eq:L}
\ee where $V'$ and $V'''$ denote respectively the first and the third
derivative of the potential (for simplicity we are restricting again
to the one dimensional case).

\section{Some Simple Examples of $\e$}
\label{subsec:Some Simple Examples of epsilon}

It can be useful to compute directly, in some simple special cases,
what is the small adimensional parameter $\e$ relevant for the \cb{}.

Note that, for quadratic potentials, {}from equation (\ref{eq:L}) it
follows that $L=+\infty$ and thus $\e=0$.\footnote{By quadratic
potential we mean $V(x)=ax^2+bx+c$ (for simplicity in one dimension).
Linear and constant potentials are included as limiting cases
respectively for $a$ only and $a$ and $b$ going to zero.} Since
$\e=0$, we have that the motion is \c{} on any scale $L_o$ chosen by
the experimenter, provided that $\l\ll L_{o}$.  (This is in complete
agreement with the standard understanding of the \cl{} for quadratic
potentials in terms of Wigner function, Feynman path integrals or Weyl
quantization \cite{robert}.)

A more interesting example is the case of the Coulomb
potential \be V(r)=\frac{qq'}{r}.
\label{eq:coulomb}
\ee
We find that  $L$ is
\be
 L\simeq r
\label{eq:Lcoulomb}
\ee If we consider the bound states of the hydrogen atom (with small
spread in energy as it is the case, e.g., for coherent states), $L$
becomes of the order of the Bohr radius $a_0$.  Thus, \be \e\simeq
\frac{\l}{a_0}\simeq \frac{1}{n},
\label{eq:slls}
\ee where $n$ is the principal quantum number.   For scattering states,
$L\simeq r$, where $r$ is simply the distance {}from the scattering
center.  This means that the scale on which the motion is classical is
not fixed but it is varying.

Consider now the case of the Yukawa potential
\be
V(r)=\frac{\ex^{-\mu r}}{r}.
\label{eq:yp}
\ee The scale of variation of this potential, according to definition
(\ref{eq:L}), is \be L\simeq\sqrt{\frac{(\mu r+1)r^2}{\mu^3r^3}}
\label{eq:ypk}
\ee For large distances, i.e. $r\ri +\infty$ (scattering states), we
have $L\simeq \frac{1}{\mu}$, the range of the potential; for small
distances, i.e. $r\ri 0$, we have $L\simeq r$, like in the Coulomb
case.

\section{Remarks and Perspectives }

We would like to underline that, while in standard \QM{} the emergence
of classicality is always connected to the permanence of a narrow
\wf{} during the motion, what arises {}from the above discussion is
that the crucial feature of the \cl{} is the formation of very spread
out \wf{}: the \LPW{}.  Only in the framework of \BM{}, given that we
have also configurations and not only the \wf{}, we can explain the
emergence of the \cb{} in a coherent way for delocalized wave
functions.

Moreover, {}from the very notion of \LPW{} it follows that the \LPW{}
can be thought as composed by a sum of \virt{} \vps{} with a definite
local \wl{} (for further details, see \cite{allori0} and
\cite{allori2}).  Since in \BM{} each particle has its own \tr{}, the
\LPW{} undergoes a sort of collapse in the following sense: not all
the \wf{} is relevant for the dynamics but there is an {\it effective}
guiding \vp{} for the particle (which is the part of the \wf{} in a
local neighborhood of the \tr{} at the time $t$).  This fact is the
key to understand the emergence of the \cl{} in \BM{} for spread out
wave functions, by applying one of the simplest argument used in
standard \QM{}, i.e. the Ehrenfest theorem.  While a detailed
explanation of how this comes about will be given in \cite{allori1}
and \cite{allori2}, we want to stress here that this argument provides
a general support for the emergence of the \cw{} whenever $\l\ll L$
and supplies with the precise notion of the scale of variation of the
potential $L$ given by equation (\ref{eq:L}).

Another delicate issue associated to the derivation of the \cl{} is
the following: as soon as there is a potential, caustics appear and
our analysis of sections \ref{subsec:Quasi Classical Wave Functions}
and \ref{subsec:Slowly Varying Potentials} for times larger than the
``first caustic time'' breaks down.  It turns out, however, that
caustics are indeed not a problem because they arise in the highly
idealized model we have considered here: we have in fact neglected the
term $H_{\rm int}$ in (\ref{eq:hamex}) describing the interaction
between the \com{} $x$ of the body and the relative coordinates $y$,
as well as any perturbation due to the unavoidable interaction of the
body with the external \env{}.  These interactions produce {\it
entanglement} between the \com{} $x$ of the system and the other
degrees of freedom $y$ (where now $y$ includes both the relative
coordinates and the degrees of freedom of the \env{}).  Taking into
account these interactions is what nowadays people call \deco{} (see,
e.g., \cite{libro} and reference therein), which however is nothing
but an {\it effective} description of all the effects that cannot be
described by the \extp{} acting on the \com{} $x$.  The role of the
external \env{} is to suppress the interference produced by the
presence of caustics \cite{allori0}.

Finally, for a complete understanding of the \cl{}, there is further
difficulty to consider: even for very small perturbations due to the
interaction with the \env{}, \sev{} for very narrow \wf{} of the
\com{} is quickly destroyed.  To solve this problem is not easy, but
the key to overcome this difficulty is to observe that the emergence
of classicality should be associated with the production of \LPWs{} in
the \com{} coordinate $x$ of the body, with weak dependence on the
other \dof{}{} $y$ (see \cite{allori0} and \cite{allori2}).

\section{Acknowledgments}

This work was financially supported in part by the INFN. Part of the
work has grown and has been developed in the IHES, the Mathematisches
Institut der Universit\"at M\"unchen and the Department of Mathematics
of the University of Rutgers.  The hospitality of these institutions
is gratefully acknowledged.  Finally, we thank Detlef D\"urr and
Shelly Goldstein for many helpful discussions, enlightening (and
crucial) suggestions and their involvement in a common project on the
derivation of the classical limit.

 \end{document}